\documentclass{jkas}
\usepackage{float} 

\def\year{2025} 
\def\volume{---} 
\def\issue{---} 
\def\beginpage{1} 
\def\received{---} 
\def\accepted{---} 
\def\published{---} 
\date{Received \received; Accepted \accepted; Published \published}

\title{%
Water Snowline in Young Stellar Objects with Various Density Structures Using Radiative Transfer Models
}

\author[1]{Young-Jun Kim}{0009-0004-7886-9029}
\author[1,2,$\star$]{Jeong-Eun Lee}{0000-0003-3119-2087}
\author[1,3]{Giseon Baek}{0000-0002-2814-1978}
\author[4]{Seokho Lee}{0000-0002-0226-9295}

\affil[1]{Department of Physics and Astronomy, Seoul National University, 1 Gwanak-ro, Gwanak-gu, Seoul 08826, Republic of Korea}
\affil[2]{SNU Astronomy Research Center, Seoul National University, 1 Gwanak-ro, Gwanak-gu, Seoul 08826, Republic of Korea}
\affil[3]{Research Institute of Basic Sciences, Seoul National University, Seoul 08826, Republic of Korea}
\affil[4]{Korea Astronomy and Space Science Institute, 776 Daedeok-daero, Yuseong-gu, Daejeon 34055, Republic of Korea}

\def\corrauthor{%
Jeong-Eun Lee, \email{lee.jeongeun@snu.ac.kr}
}

\def\runningauthor{%
Kim Y.-J. et al. 2025
}

\def\runningtitle{%
Water Snowline in Young Stellar Objects with Various Density Structures Using Radiative Transfer Models
}

\def\keywords{%
astrophysics --- stars: protostars --- protoplanetary discs --- radiative transfer
}

\def\abstracttext{%
Tracing the water snowline in low-mass young stellar objects (YSOs) is important because dust grain growth is promoted and the chemical composition varies at the water snowline, which influences planet formation and its properties. In protostellar envelopes, the water snowline can be estimated as a function of luminosity using a relation derived from radiative transfer models, and these predictions are consistent with observations. However, accurately estimating the water snowline in protoplanetary disks requires new relations that account for the disk structure. We present the relations between luminosity and water snowline using the dust continuum radiative transfer models with various density structures. We adopt two-dimensional density structures for an envelope-only model (Model E), an envelope+disk+cavity model (Model E+D), and a protoplanetary disk model (Model PPD). The relations between the water snowline, where $T_{\rm dust}$ = 100 K, and the total luminosity, ranging 0.1–1,000 $L_{\odot}$, are well fitted by a power-law relation, $R_{snow}=a\times (L/L_{\odot})^p$ au. The factor $a$ decreases with increasing disk density, while the power index $p$ has values around 0.5 in all models. As the disk becomes denser, the water snowline forms at smaller radii even at the same luminosity, since dense dust hinders photon propagation. We also explore the effect of viscous heating on the water snowline. In Model PPD with viscous heating, the water snowline shifts outward by a few au up to 15 au, increasing the factor $a$ and decreasing the power index $p$.  In Model E+D with lower disk mass, the effect of viscous heating is negligible, indicating that the disk mass controls the effect. The discrepancy between our models and direct observations provides insights into the recent outburst event and the presence of a disk structure in low-mass YSOs.
}

\begin{document}
\jkashead 

\begin{figure*}[t]
\centering
\includegraphics[width=\textwidth]{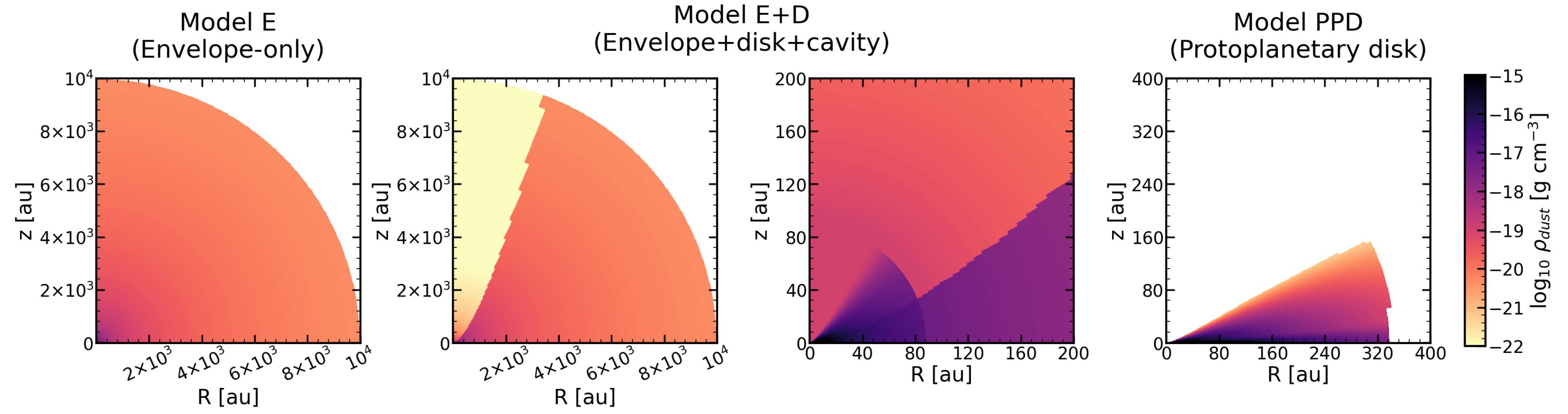}
\caption{The dust mass density distributions of low-mass YSOs. The first column shows an envelope-only model (Model E). The second column shows an envelope + disk + cavity model (Model E+D), and the third column shows a zoomed-in view of Model E+D within 200 au. The fourth column shows a protoplanetary disk model (Model PPD).
\label{fig:fig01}}
\end{figure*}
\section{Introduction}
\label{sec:section1}
Tracing the water snowline, where the phase transition of water molecules occurs from gas to ice, provides key information to understand the physical and chemical evolution of low-mass young stellar objects (YSOs). In particular, the water snowline in the midplane of a protoplanetary disk is directly related to planet formation. The growth of dust grains is enhanced beyond the water snowline, allowing planet formation with km-sized planetesimals (\citealt{Stevenson1988, Saito2011, Gundlach2014, Zhang2015, Schoonenberg2017a, Schoonenberg2017b, Houge2024}). 

In addition, the chemical composition of the planets, often expressed as the C/O ratio, is determined based on the location of the water snowline (\citealt{Oberg2011, Eistrup2016, Mordasini2016, Eistrup2018}). Furthermore, complex organic molecules (COMs) trapped within water ice are released into the gas phase along with the water when dust grains are heated (\citealt{Herbst2009, vanDishoeck2013}). Thus, the water snowline in the protoplanetary disk plays a crucial role in shaping the chemical diversity during planet formation.

The water snowline is traced by directly observing water molecules and other chemically connected molecules. For example, in the protostellar envelope, spatially resolved water molecular emissions have been revealed through (sub)millimeter observations (\citealt{Jorgensen2010, Persson2012, Persson2013, Persson2014, Persson2016, Jensen2019, Jensen2021}). Alternatively, HCO$^{+}$, which shows an anti-correlation with H$_{2}$O (\citealt{Phillips1992, Bergin1998}), has indirectly identified the water snowline in the protostellar envelope (\citealt{vant2018a, vant2022}).

However, in protoplanetary disks the dense disk structure confines the water snowline, with a higher sublimation temperature of 100-200 K, to less than a few au (\citealt{Hayashi1981, DAlessio2001, Kennedy2008}). This makes it difficult to directly resolve the water snowline through observations of protoplanetary disks in the quiescent phase, requiring a higher angular resolution (\citealt{Kristensen2016, Carr2018, Notsu2019, Bosman2021, Facchini2024, Guerra2024}). However, even in a disk, an enhanced luminosity resulting from an accretion outburst can extend the water snowline to tens of au (\citealt{Harsono2015,Lee2019}).

The disk-dominant Class I/II protostar V883 Ori, an FU Orionis type object, is undergoing an outburst with a total luminosity of 218-647 $L_{\odot}$ (\citealt{Strom1993, Furlan2016, Liu2022}), allowing the water snowline to be resolved in observations. \citet{Cieza2016} estimated the water snowline of  42 au based on 1.3mm dust continuum observation. The location of the water snowline was also estimated indirectly by observations of COMs such as methanol as well as HCO$^{+}$ (\citealt{vant2018b, Lee2019, Leemker2021}). Recently, the water molecular emissions were detected in V883 Ori, and the largest water sublimation radius was 120–160 au, which is inferred to originate from the disk surface  (\citealt{Tobin2023, Lee2024}). These studies estimated the water snowline as $\sim$80 au on the midplane.

The water snowline is explored not only through observations but also using radiative transfer models. \citet{Bisschop2007} investigated the water sublimation radius ($T_{\text{dust}}$ = 100 K) in the high-mass protostellar envelope using a 1D dust continuum radiative transfer model, demonstrating that the water snowline follows the relation in Equation \ref{eq:bisschop} for a luminosity range of $10^{4}$–$10^{6}$ $L_{\odot}$. 
\begin{equation}
\label{eq:bisschop}
 R(T_{\rm dust}=100 K) =15.4\times\sqrt{L/L_{\odot}} \quad \text{[au]}
\end{equation}
Furthermore, \citet{vant2022} showed that this relation is also applicable to low-mass protostellar envelopes at a luminosity range of 0.5–50 $L_{\odot}$ and can be compared with the water snowlines detected by the water emissions.

However, the water snowline in the protoplanetary (e.g., V883 Ori) disk predicted by an envelope-only model can lead to misinterpretations of the central luminosity. Equation \ref{eq:bisschop} with the current luminosity of V883 Ori significantly overestimates the snowline to 230–360 au, compared to the observed value (42–80 au). Furthermore, in V883 Ori, which has a high mass accretion rate, viscous torque shearing actively heats the disk midplane \citep{Liu2022}, causing the water snowline to shift outward by $\sim$10 au \citep{Alarcón2024}. 

Since the envelope-only model from \citet{Bisschop2007} cannot accurately predict the location of the water snowline in protoplanetary disks, tracing the snowline using radiative transfer models requires the application of density structures and energy sources appropriate to the system. This study explores the water snowline as traced by the dust temperature obtained from radiative transfer models  with various density structures and energy sources. The models of low-mass YSOs are described in Section \ref{sec:method2}. In Section \ref{sec:result}, we derive the relations between luminosity and water snowline, obtained from the models. The capacity of tracing the water snowline with our relation is discussed in Section \ref{sec:discussion}. Lastly, in Section \ref{sec:conclusion}, we present the conclusion of this study.


\section{Radiative Transfer Model}
\label{sec:method2}
To explore the effect of the density structure on the water snowline, we first construct two-dimensional density structures for low-mass YSOs: envelope-only model (Model E), envelope + disk + outflow cavity model (Model E+D), and protoplanetary disk model (Model PPD). We create the gas density distributions and define the dust density distributions using a gas-to-dust ratio of 100:1. The dust density distributions of three models are presented in Figure \ref{fig:fig01} and the parameters for each model are provided in Table \ref{tab:table_class0} – \ref{tab:table_class2}.

\begin{figure}
\centering
\includegraphics[width=0.45\textwidth]{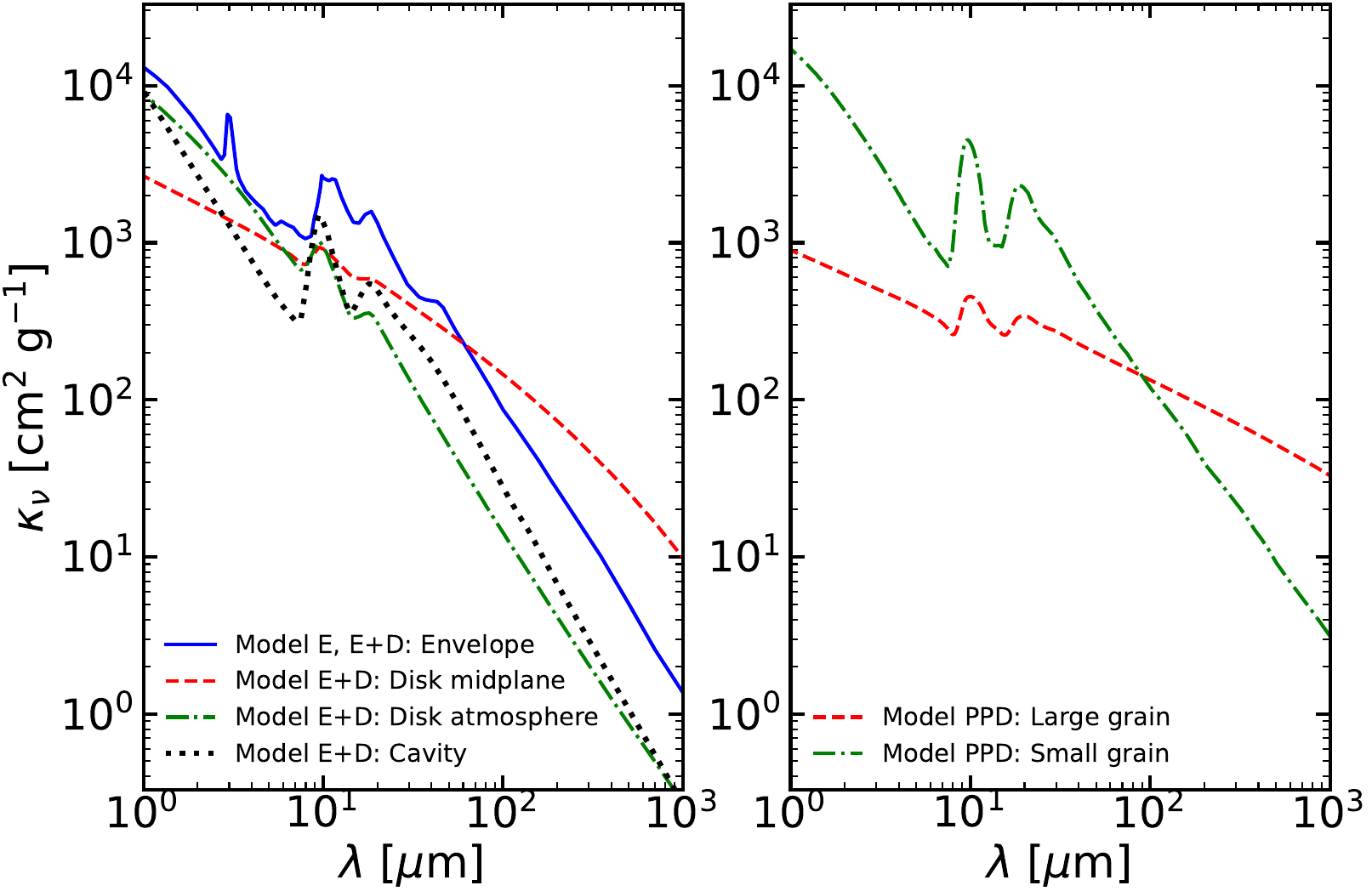}
\caption{Total dust opacity $\kappa_{\nu}$ with the wavelength. The dust opacity profiles for different structures in Model E and E+D are adopted from \citet{Baek2020} \textit{(left)}. The dust opacity profiles for two different grain size populations in Model PPD are adopted from \citet{DAlessio2006} \textit{(right)}.\label{fig:fig02}}
\end{figure}

For Model E and E+D, the spatial grids are defined in a spherical coordinate system with the grid numbers of ($n_{r}, n_{\theta}, n_{\phi}$) = (200, 100, 1). In the radial direction, a logarithmic scale is adopted. However, because snowlines form at large radii under high luminosities, $n_{r}$ increases from 200 to 800 depending on the luminosity to accurately trace the snowline in larger radii: 
\begin{itemize}
  \item $L_{tot}$ = 0.1–1 $L_{\odot}$: $n_{r}$ = 200
  \item $L_{tot}$ = 1–10 $L_{\odot}$: $n_{r}$ = 400
  \item $L_{tot}$ = 10–100 $L_{\odot}$: $n_{r}$ = 600
  \item $L_{tot}$ = 100–1,000 $L_{\odot}$: $n_{r}$ = 800
\end{itemize}
For Model PPD, we fix the radial grid cells to $n_{r}$ = 1,000 across all luminosities to prevent calculation uncertainty caused by the optically thick disk. The inner radii of all models are adjusted to satisfy $T_{\rm dust}$ = 1,200 K, considering the thermal destruction of dust \citep{Mac2019a}. In the polar direction, we employ a non-uniform grid with a power-law stretching toward the midplane to trace the snowline accurately.

Next, we apply different dust opacity profiles to the individual components of each model to reflect the grain size distribution in each structure. Last, we use RADMC-3D \citep{Dullemond2012} to calculate the dust temperature distributions for these models, considering various energy sources (e.g, passive heating, active heating) across a total luminosity range of $L_{\rm tot}$ = 0.1–1,000 $L_{\odot}$. RADMC-3D is a code package that calculates the radiative transfer process using the Monte Carlo Radiative Transfer Simulation method \citep{Bjorkman2001}. 

To calculate the dust temperature in the models, we adopt a photon number of $N_{phot}$ = 10$^{7}$ for Model E and E+D. As Model PPD has an optically thick disk, we adopt $N_{phot}$= 10$^{8}$ to calculate the dust temperature accurately. In models with viscous heating, since an additional heat source is included in each grid cell (Figure \ref{fig:fig03}), we confirm that the dust temperature distribution with $N_{phot}$= 10$^{6}$ of the photon packet is sufficiently continuous. These numbers are selected to optimize the accuracy and computational efficiency of the dust temperature calculations.

We note that each model is not representative, but selected examples at different evolutionary stages. The actual density structures and properties of the components (e.g., disk mass) vary between systems; therefore, the actual water snowlines differ depending on the protostellar system.


\subsection{Model E: Envelope-Only}
\label{sec:method_Class0}
\subsubsection{Density structure}
In this envelope-only model, we adopt a simple spherically symmetric power-law density profile as
\begin{equation}
\label{eq:envelope}
 \rho_{\rm env}(r) = \rho_{\rm env,0}\left(\frac{r}{R_{\rm env,in}}\right)^{-p}.
\end{equation}
$\rho_{\rm env,0}$ is the gas mass volume density at $R_{\rm env,in}$ and this value is determined by the envelope mass $M_{\rm env}$. $R_{\rm env,in}$ is the inner radius of the envelope, and we vary the envelope mass in the range of $M_{\rm env}$ = 0.5–6.5 $M_{\odot}$. $r$ is the radius in spherical coordinates, and $p$ is the power-law index of the radial density profile. We assume that the infall is free-fall, with $p=1.5$. For the fiducial model, we adopt $M_{\rm env}$ = 6.5 $M_{\odot}$, which has the same envelope density structure as Model E+D. The values of the model parameters are given in Table \ref{tab:table_class0}.
 
\subsubsection{Dust property}
As Model E consists of only the thick envelope, we adopt the opacity of the OH5 dust, which coagulates at a density of $10^{6}$\,cm$^{-3}$ for $10^{5}$\,yr and has a thin ice mantle, taken from the 5th column of Table 1 in \citet{OH5_1994}. 

\begin{table}
\caption{Parameters of Model E\label{tab:table_class0}}
\centering
\scriptsize
\resizebox{\columnwidth}{!}{%
\begin{tabular}{lrr}
\toprule
Parameter                         & Description                 & Values \\
\midrule
$L_{\rm tot}$   [$L_{\odot}$]&  Total luminosity           &  0.1–1,000, (\textbf{6})\\
$T_{*}$   [K]&  Stellar temperature        &  4,000      \\ \addlinespace
$M_{\rm env}$   [$M_{\odot}$]&  Envelope gas mass&  0.5, 1, 2, 3, 5, (\textbf{6.5})      \\
$R_{\rm env,in}$   [au]&  Envelope inner radius
&  Varying, (\textbf{1.61})\\
$R_{\rm env,out}$   [au]&  Envelope outer radius
&  10,000\\
$\rho_{\rm env,0}$   [$g/cm^{3}$]&  Envelope gas density at $R_{\rm e,in}$&  Varying, (\textbf{2.25$\times 10^{-13}$})\\
$p$                               &  Envelope radial density power-law index&  1.5      \\
\bottomrule
\end{tabular}%
}
\tabnote{\textit{Note:} For the luminosity-dependent values, those in bold indicate the fiducial model parameters.}
\end{table}

\subsubsection{Energy source}
\label{sec:modelE_energy}
For the Model E, we assume that the energy source includes the stellar and accretion luminosity ($L_{\rm tot}$=$L_{*}$+$L_{acc}$) without external heating. Estimating the stellar properties in deeply embedded protostellar envelopes is a challenging task. In addition, since the main purpose of our models is to trace the water snowline in various density structures, we adopt the simple assumption for stellar properties. \citet{Whitney2003b} and \citet{Furlan2016} used the stellar properties of a stellar atmosphere model with a temperature of 4,000 K \citep{Kurucz1994}, which represents a typical T Tauri star, to model the low-mass YSOs in various density structures.

Following this approach, we assume that the energy source emits blackbody radiation of  $T_{*}$ = 4,000 K. In the case of blackbody radiation, RADMC-3D adopts the radius of irradiative (passive) heating energy source ($R_{\rm irr}$) to generate the energy source with the following equation;
\begin{equation}
\label{eq:input_radius}
 R_{ \rm irr} = R_{\odot} \times \sqrt{L_{\rm tot}/L_{\odot}} \times (T_{*}/T_{\odot})^{2}.
\end{equation}
Thus, we control the energy source with the value of the total luminosity and trace the water snowline in the luminosity range of $L_{\rm tot}$ = 0.1–1,000 $L_{\odot}$.

\subsection{Model E+D: Envelope + Disk + Cavity}
\label{sec:method_Class1}
\subsubsection{Density structure}
Model E+D comprises an envelope, a disk, and a bipolar outflow cavity. To create these structures, we adopt the best-fit model parameters for EC 53 (\citealt{Baek2020}, Table 1), which is classified as a late Class 0 to early embedded Class I YSO (\citealt{Dunham2015, Lee2020}). The envelope density structure in this model also follows eq~(\ref{eq:envelope}). For the disk in Model E+D, we use the standard flared disk structure (\citealt{Shakura1973, Lynden1974, Hartmann1998}), given as
\begin{multline}
\label{eq:disk_vol}
 \rho(R,z) = \rho_{\rm disk,0} \left(1-\sqrt{\frac{R_{*}}{R}}\right) \left(\frac{R_{*}}{R}\right)^{\alpha} \\
 \times \text{exp} \left[-\frac{1}{2} \left(\frac{z}{H}\right)^{2}\right].
\end{multline}
$\rho_{disk,0}$ is the normalized constant, and the disk mass is calculated by the integral of the density structure. $R_{*}$ is the star radius and $\alpha$ is the power-law index that governs the radial density distribution. $R$ and $z$ are the radius and zenith in cylindrical coordinates. The disk scale height $H$ (in au) is defined as $H = H_{100}\times(R/100 \text{ au})^{\beta}$, where $H_{100}$ is the disk scale height at 100 au, and $\beta$ is the disk flaring power index.

At this stage, to counteract the angular momentum generated by mass accretion, bipolar outflows expel material at high velocities, creating cavities in the surrounding envelope. To represent this, we implement a curved cavity structure aligned perpendicular to the disk plane

\begin{equation}
\label{eq:cavity}
 z = c R^{d}.
\end{equation}
The constant $c$ is defined as $c = R_{\rm env,out}$/$(R_{\rm env,out}$ tan $\theta_{cav})^{1.5}$, where $R_{\rm env,out}$ is envelope outer radius and $\theta_{cav}$ is cavity opening angle. We adopt $d$ = 1.5 as the cavity shape exponent, following \citet{Baek2020}. The values of the model parameters are given in Table \ref{tab:table_class1}.

\subsubsection{Dust property}
For the dust opacity profile in each structure of Model E+D (Figure \ref{fig:fig02}, left), we follow \citet{Baek2020}; dust opacity profiles for the disk are adopted from \citet{Wood2002} and \citet{Cotera2001}, which were used to fit the observation of HH 30, including both the midplane ($n(H_{2})>10^{10}$ cm$^{-3}$) and atmosphere ($n(H_{2})<10^{10}$ cm$^{-3}$). For the envelope, we use the OH5 opacity, which was used for the envelope in Model E. Lastly, the opacity of the outflow cavity structure is based on \citet{Kim1994}, similar to the ISM in Taurus.

\subsubsection{Energy source}
Model E+D is assumed to have the energy source of passive heating from central star and mass accretion ($L_{\rm tot}$=$L_{*}$+$L_{acc}$). We follow the stellar properties of \citet{Baek2020}, as a reference of Model E+D. They also assumed the central star to be the typical T Tauri star with a stellar temperature of 4,000 K; therefore, we use the same energy source setup as in Section. \ref{sec:modelE_energy}.

However, in Section \ref{sec:discussion_2}, since we discuss the effect of viscous (active) heating in Model E+D, which requires the stellar mass and radius (Equation \ref{eq:model_ppd_lum}), we adopt the additional stellar properties from \citet{Baek2020}. They adopted a stellar mass of 0.5 $M_{\odot}$, a radius of 2.09 $R_{\odot}$, and a luminosity of 1 $L_{\odot}$. Thus, we explore the water snowline in a luminosity range of $L_{\rm tot}$ = 1–1,000 $L_{\odot}$.

\begin{table}
\caption{Parameters of Model E+D\label{tab:table_class1}}
\centering
\scriptsize 
\resizebox{\columnwidth}{!}{%
\begin{tabular}{lrr}
\toprule
Parameter                         & Description                       & Values \\
\midrule
$L_{\rm tot}$   [$L_{\odot}$]&  Total luminosity                 &  1–1,000, (\textbf{6})\\
$L_{*}$   [$L_{\odot}$]& Stellar luminosity&1.0\\
$M_{*}$   [$M_{\odot}$]&  Stellar mass                     & 0.5      \\
$R_{*}$   [$R_{\odot}$]&  Stellar radius&  2.09 \\
$T_{*}$   [K]&  Stellar temperature&  4,000\\ \addlinespace
$M_{\rm disk}$   [$M_{\odot}$]&  Disk gas mass&  0.0075      \\
$R_{\rm disk,in}$   [au]&  Disk inner radius                &   Varying, (\textbf{0.34})     \\
$R_{\rm disk,out}$   [au]&  Disk outer radius                &   90     \\
$H_{100}$ [au]&  Disk scale height at 100 au&   48     \\
$\alpha$                          &  Disk radial density exponent     &   2.5     \\
$\beta$                           &  Disk flaring power-law index     &   1.3     \\
\addlinespace
$M_{\rm env}$   [$M_{\odot}$]&  Envelope gas mass&  5.8      \\
$R_{\rm env,in}$   [au]&  Envelope inner radius&  Varying, (\textbf{0.34})\\
$R_{\rm env,out}$   [au]&  Envelope outer radius&  10,000\\
$\rho_{\rm env,0}$   [$g/cm^{3}$]&  Envelope gas density at $R_{\rm env,in}$&  Varying, (\textbf{2.31$\times 10^{-12}$})\\
$p$                               &  Envelope radial density power-law index&  1.5      \\ \addlinespace
$d$&  Cavity shape exponent            &  1.5      \\
$\theta_{\rm cav}$   [$^{\circ}$]&  Cavity opening angle             &  20      \\
$\rho_{\rm cav,in}$   [$g/cm^{3}$]&  Cavity inner gas density&  $10^{-17}$\\
$R_{\rm cav,bd}$   au&  Cavity inner boundary radius     &  100      \\
\bottomrule
\end{tabular}%
}
\tabnote{\textit{Note:} For the luminosity-dependent values, those in bold indicate the fiducial model parameters.}
\end{table}


\subsection{Model PPD: Protoplanetary Disk}
\label{sec:method_Class2}
\subsubsection{Density structure}
Model PPD represents a disk-only system. To model this environment, we adopt the parameters for V883 Ori (\citealt{Leemker2021}, Table D.1.), which is classified as a late Class I to early Class II YSO \citep{Furlan2016}. Additionally, V883 Ori is the only disk source with a resolved water snowline. The values of the model parameters are given in Table \ref{tab:table_class2}. In V883 Ori, the water snowline has been resolved through observations (\citealt{Cieza2016, vant2018b, Lee2019, Tobin2023, Lee2024}) and radiative transfer models have been suggested (\citealt{Cieza2018} ; \citealt{Leemker2021}), making it appropriate as a reference.

The protoplanetary disk models the vertical distribution of two dust grain populations following the approach of \citet{DAlessio2006}. Although Model E+D adopts the midplane and atmosphere structures in the disk, these regions are simply divided by the gas density of $n(H_{2})$ = $10^{10}$ cm$^{-3}$. However, in disk-only Model PPD, we clearly present dust settling by applying the scale height factor and the mass fraction factor to the large grain population, as described by \citet{Simon2013}.
The density distributions for large and small grains are given by the following equations: 
\begin{equation}
\label{eq:rhod_l}
 \rho_{\rm  d, large}(R,z) = \frac{f_{lg} \Sigma}{\sqrt{2\pi} R \chi h} \text{exp} \left[-\frac{1}{2} \left(\frac{z}{R \chi h}\right)^{2}\right],
\end{equation}

\begin{equation}
\label{eq:rhod_s}
 \rho_{\rm d,small}(R,z) = \frac{(1-f_{lg}) \Sigma}{\sqrt{2\pi} R h} \text{exp} \left[-\frac{1}{2} \left(\frac{z}{R h}\right)^{2}\right].
\end{equation}
$f_{lg}$ and $\chi$ are the mass fraction and the scale height factor for large grain dust, respectively. We assume that the vertical density distribution follows a Gaussian profile based on the disk scale height $h = h_{c} (R/R_{c})^{\psi}$. The gas surface density $\Sigma$ of the flared disk (\citealt{Shakura1973, Lynden1974, Hartmann1998}) is given as
\begin{equation}
\label{eq:disk_surf}
 \Sigma(R) = \Sigma_{c} \left(\frac{R}{R_{c}}\right)^{-\gamma} \text{exp} \left[-\left(\frac{R}{R_{c}}\right)^{2-\gamma}\right].
\end{equation}
$\Sigma_{c}$ is the normalized surface density of a disk with mass $M_{d}$ at a characteristic radius $R_{c}$, and is given by
\begin{equation}
\label{eq:sig_c}
 \Sigma_{c} = (2-\gamma) \frac{M_{d}}{2 \pi R_{c}^{2}}.
\end{equation}

\subsubsection{Dust property}
For the dust opacity profile in each structure of Model PPD (Figure \ref{fig:fig02}, right), we follow \citet{Simon2013}. The dust opacity profiles for the disk are modeled with two size distributions; the large grain population ($a$ = 5nm–1mm) settles in the midplane, and the small grain population ($a$ = 5nm–1$\mu$ m) comprises the atmosphere of the disk. Both populations follow the MRN size distribution, which is proportional to $a^{-3.5}$ \citep{MRN1977}.

\begin{figure}
\includegraphics[width=0.45\textwidth]{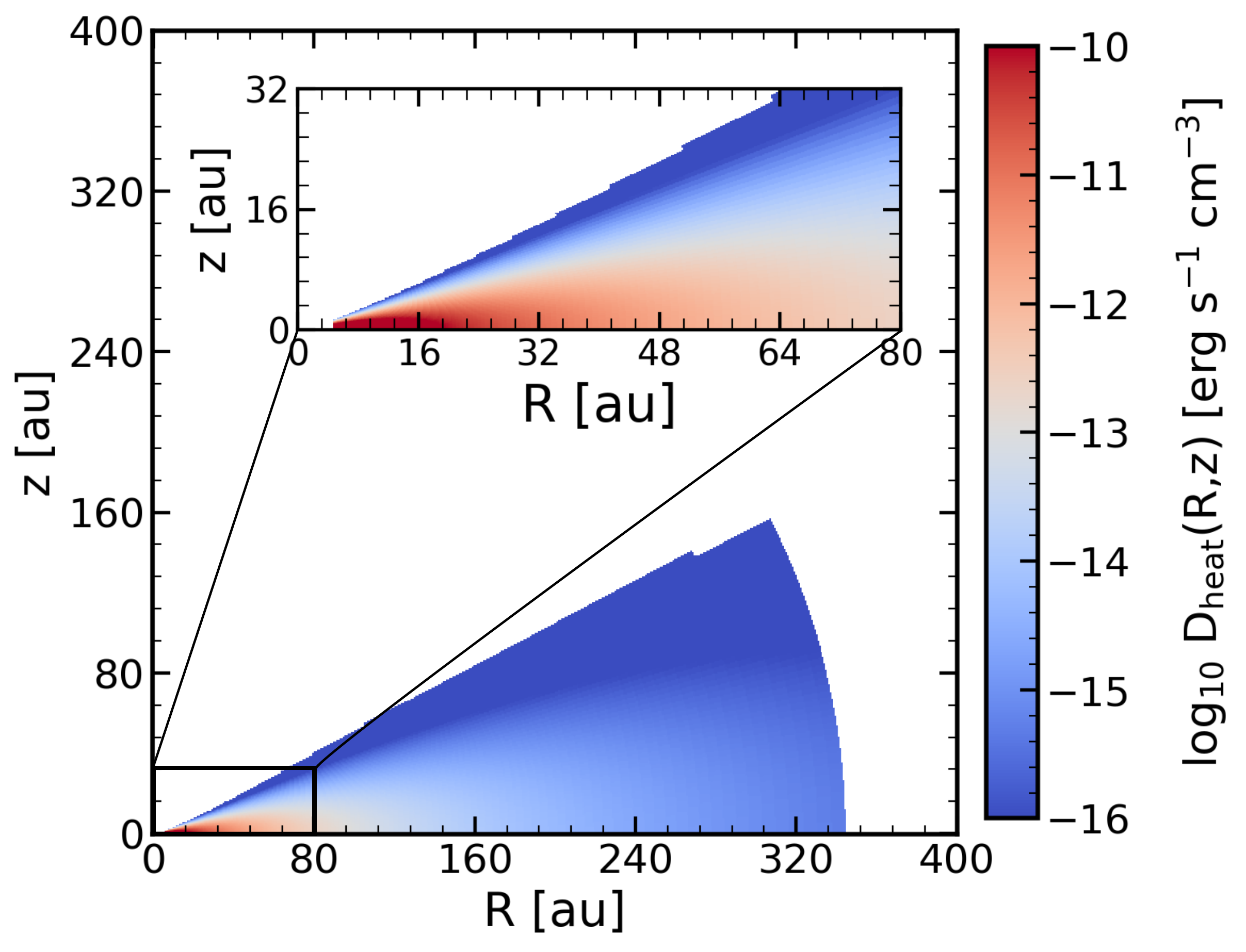}
\caption{The 2-dimensional viscous heat dissipation distribution of Model PPD, where $\dot{M} = 5.32 \times 10^{-5} M_{\odot} yr^{-1}$.  The inset in the upper panel zooms in on the inner 80 au$\times$32 au. \label{fig:fig03}}
\end{figure}

\begin{table}
\caption{Parameters of Model PPD\label{tab:table_class2}}
\centering
\scriptsize 
\begin{tabular}{lrr}
\toprule
Parameter                               & Description                       & Values \\
\midrule
$L_{\rm tot}$   [$L_{\odot}$]&  Total luminosity                 &  10–1,000, (\textbf{400})\\
$L_{*}$   [$L_{\odot}$]&  Stellar luminosity&  6.0\\
$M_{*}$   [$M_{\odot}$]
&  Stellar mass                     & 1.22$^{\rm \dag}$      \\
$R_{*}$   [$R_{\odot}$]&  Stellar radius&  5.1\\
$T_{*}$   [K]&  Stellar temperature&  10,000\\
$\dot M$   [$M_{\odot}/yr$]&  Mass accretion rate              &   Varying, (\textbf{5.32$\times 10^{-5}$}$^{\rm \dag}$)      \\ \addlinespace
$M_{\rm disk}$   [$M_{\odot}$]&  Disk gas mass&  0.23 $^{\rm \dag}$     \\
$R_{\rm disk,in}$   [au]&  Disk inner radius                &   Varying, (\textbf{4.61} $^{\rm \dag}$)     \\
$R_{\rm disk,out}$   [au]&  Disk outer radius                &   338 $^{\rm \dag}$     \\
$R_{c}$   [au]&  Characteristic radius            &   75     \\
$\Sigma_{c}$   [$g/cm^{2}$]&  Disk gas surface density&   55.22 $^{\rm \dag}$\\
$\gamma$                                &  Radial density power-law index&   1     \\
$h_{c}$                                 &  Disk scale height                &  0.1      \\
$\psi$                                  &  Disk flaring power-law index&  0.25      \\
$\chi$                                  &  Large grain scale height factor&  0.2      \\
$f_{lg}$&  Large grain mass fraction&  0.9      \\
\bottomrule
\end{tabular}
\tabnote{\textit{Note:} For the luminosity-dependent values, those in bold indicate the fiducial model parameters. \\
$^{\rm \dag}$: The stellar mass, disk gas mass, outer radius, and the disk gas surface density are changed due to distance correction according to \citet{Lee2019}. The mass accretion
rate is derived by solving the equation form $\dot M$, assuming a total luminosity of $L_{\rm tot} = L_{*}+L_{\rm acc}+L_{\rm vis}$ (Equation \ref{eq:model_ppd_lum}).}
\end{table}

\subsubsection{Energy source}
The viscous torque inside the disk becomes important in the disk-only Model PPD. Inside the disk, the difference in torque with radius causes shearing, making the disk act as its own heating source. We include viscous heating in Model PPD when adopting the energy source. The viscosity is expressed from the steady state of the Keplerian disk (\citealt{Pringle1981}; \citealt{Frank2002})
\begin{equation}
\label{eq:viscosity}
 \nu \Sigma = \frac{\dot M}{3 \pi},
\end{equation}
$\dot M$ is the mass accretion rate and $\Sigma$ is the gas surface density. The torque exerted per unit length due to the viscous disk, $G(R,z)$ can be written as
\begin{equation}
\label{eq:viscous torque}
 G(R,z) = 2 \pi R \nu \rho R^{3} \Omega'.
\end{equation}
We derive $\rho(R,z)$ assuming a vertical Gaussian distribution of the surface density with scale height $h$. $\Omega = \sqrt{GM_{*}/R^{3}}$ is the angular velocity of the disk, and $\Omega' = d\Omega/dR$. Next, the energy loss per unit volume due to heat dissipation from the viscous torque $D_{heat}(R,z)$ is expressed as
\begin{equation}
\label{eq:heat dissipation}
 D_{heat}(R,z) = \frac{G(R,z) \Omega'}{2 \pi R} = \frac{3 \dot M}{4 \pi} \frac{\Omega^{2} \rho}{\Sigma}.
\end{equation}
Then we adopt the viscous heating luminosity by multiplying the heat dissipation (Figure \ref{fig:fig03}) with the volume of each spatial grid cell ($V_{\rm grid}$). Including viscous heating, we assume the total luminosity to be
\begin{equation}
\label{eq:model_ppd_lum}
\begin{split}
L_{\rm tot} &= L_{*} + L_{acc} + L_{vis} \\
&= L_{*} + \frac{G M_{*} \dot{M}}{R_{*}}
   + \sum_{\rm grid} D_{\rm heat} \, V_{\rm grid}.
\end{split}
\end{equation}

\begin{figure*}[t]
\centering
\includegraphics[width=\textwidth]{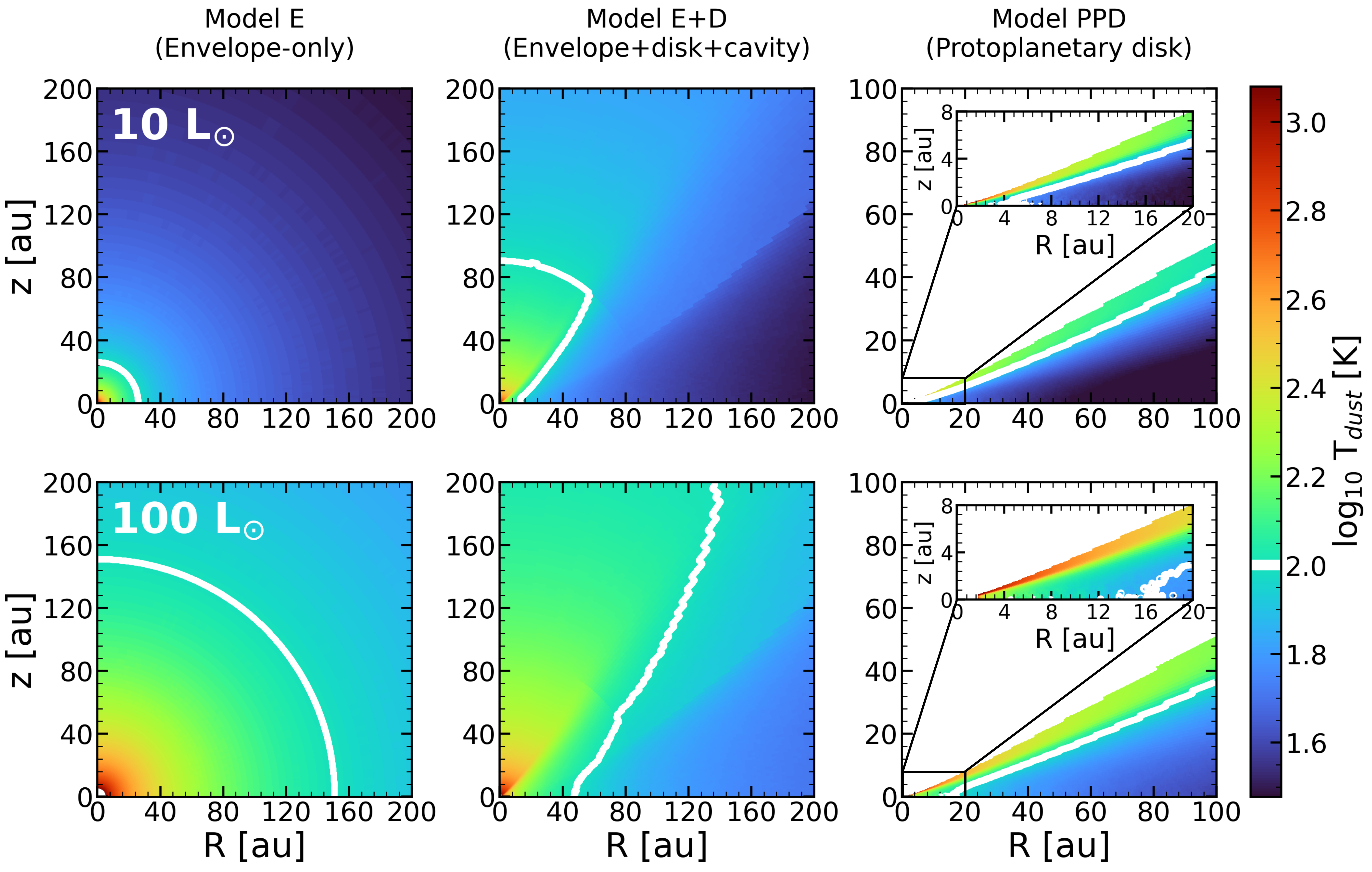}
\caption{The 2-dimensional dust temperature distributions without viscous heating are shown for 10 $L_{\odot}$ (\textit{top}) and 100 $L_{\odot}$ (\textit{bottom}); Model E \textit{(left)}, Model E+D \textit{(middle)}, and Model PPD \textit{(right)} are presented. The white solid line indicates the isothermal contour of $T_{\rm dust}$ = 100 K. The upper zoom-in panels of Model PPD show the water snowline inside the box of 20au$\times$8au. \label{fig:fig04}}
\end{figure*}

We adopt the stellar parameters of V883 Ori. \citet{Cieza2016} estimated the stellar mass to be 1.3 $M_{\odot}$ based on Keplerian rotation. They estimated a stellar luminosity of  6 $L_{\odot}$ using a pre-main sequence stellar model \citep{Siess2000}. \citet{Leemker2021} used a stellar radius of 5.1 $R_{\odot}$ and a temperature of 10,000 K to reproduce the accretion luminosity of 400 $L_{\odot}$. In this study, we adopt these stellar parameters, with the stellar mass updated to 1.22 $M_{\odot}$ based on the distance correction by \citet{Lee2019}. Since we set the stellar luminosity to be 6 $L_{\odot}$, we explore the water snowline in a luminosity range of $L_{\rm tot}$ = 10–1,000 $L_{\odot}$ for Model PPD, including irradiative (passive) and viscous (active) heating. The model parameters are provided in Table \ref{tab:table_class2}.

\begin{figure*}[t]
\centering
\includegraphics[width=\textwidth]{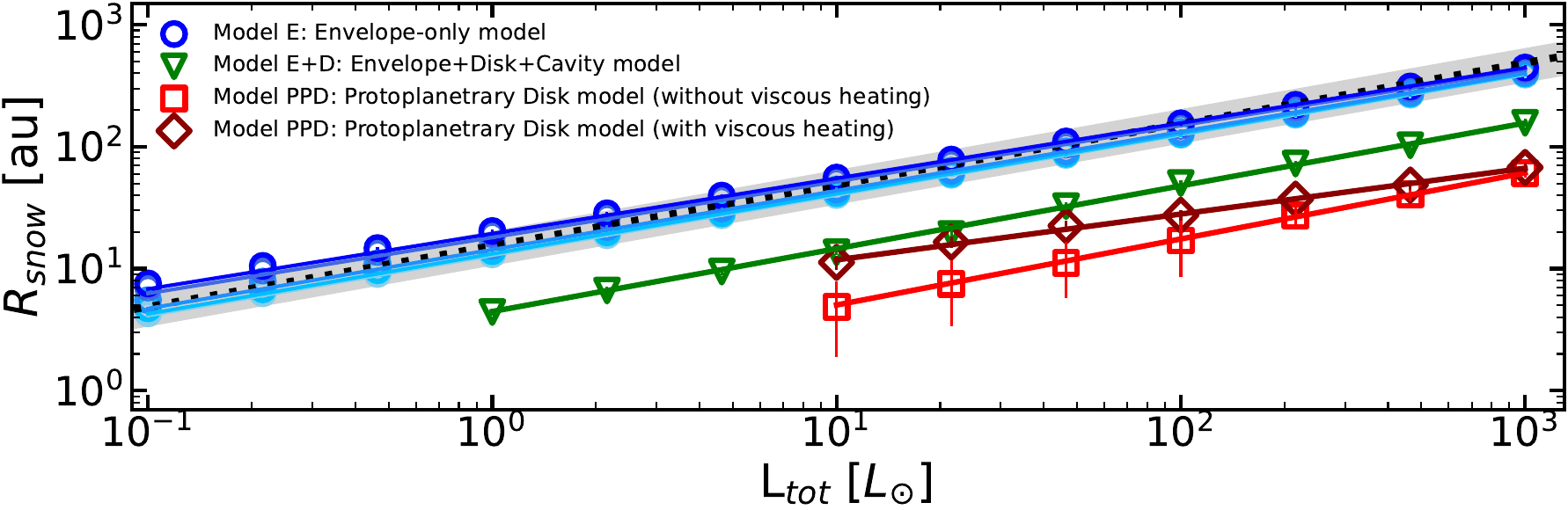}
\caption{The relation between luminosity and the water snowline in three different density structures. Each symbol represents the location of the water snowline in the model for each luminosity, and the colored solid lines represent the fitting relations of $R_{\rm snow}=a\times (L/L_{\odot})^p$ au for each model. Bluish colors and circle symbols represent Model E. Colors from lightest to darkest blue represent models with envelope masses of 0.5, 1.0, 2.0, 5.0, and 6.5 $M_{\odot}$, respectively. Green and triangle symbols represent Model E+D. Reddish colors represent Model PPD. Red and square symbols represent models without viscous heating, while dark red and diamond symbols represent models with viscous heating. The black dashed line corresponds to the relation from \citet{Bisschop2007} and the gray shade area represents a 30\% uncertainty range suggested by \citet{vant2022}. \label{fig:fig05}}
\end{figure*}

\section{Result}
\label{sec:result}
\begin{figure}[t]
\centering
\includegraphics[width=0.45\textwidth]{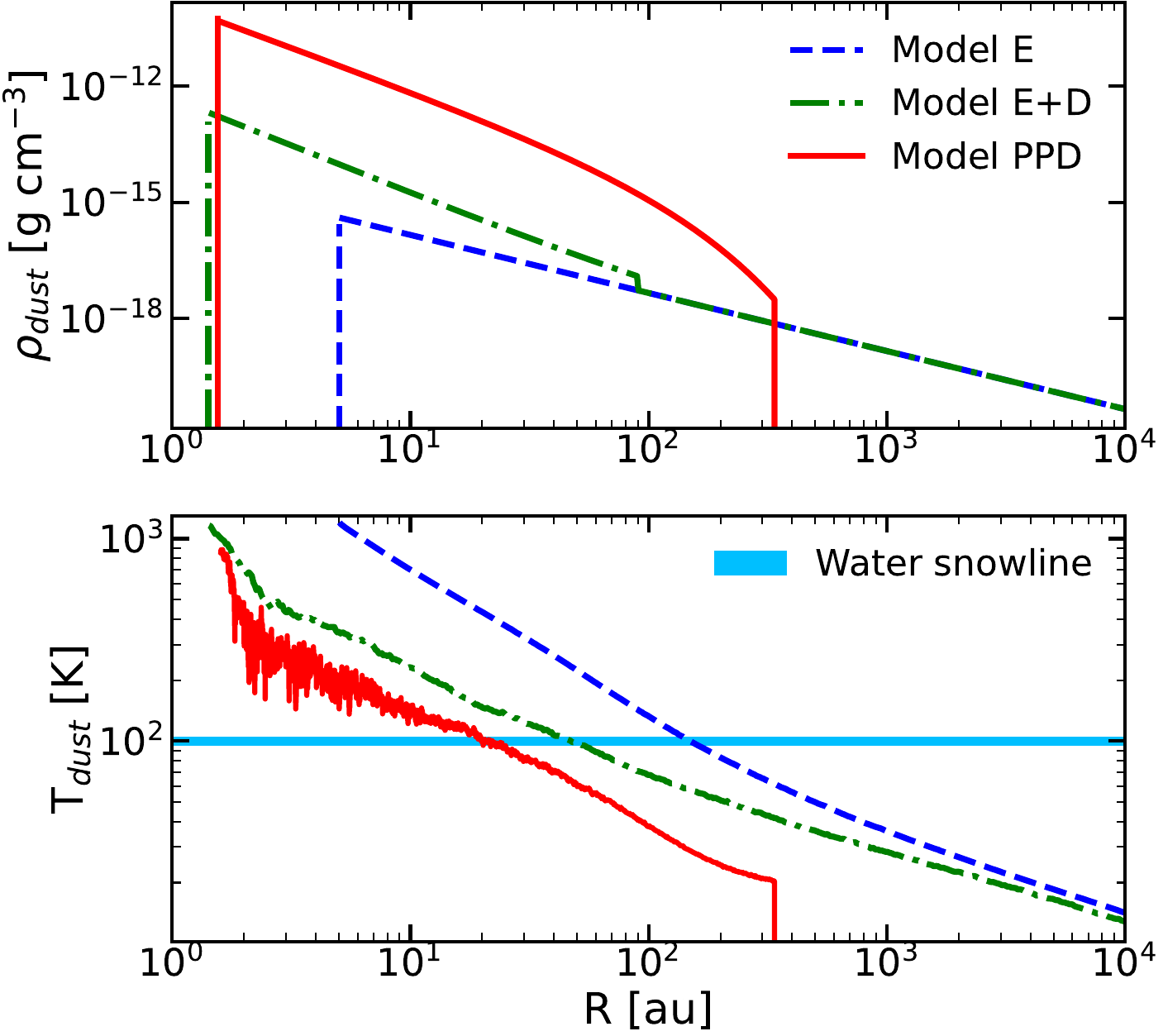}
\caption{The dust density (\textit{top}) and temperature distributions (\textit{bottom}) in the midplane. The dashed, dash-dotted, and solid lines represent Model E, E+D, and PPD  (without viscous heating), respectively. The models have $L_{\rm tot}$= 100 $L_{\odot}$, and viscous heating is not included. The skyblue line denotes $T_{\rm dust}$ = 100 K, which defines the water snowline. \label{fig:fig06}}
\end{figure}
In this study, we define the water snowline as the distance from the protostar in the midplane where the dust temperature is 100 K. Although the water sublimation temperature could be 160-200 K due to the higher pressure in the disk structure (\citealt{Fraser2001}; \citealt{Podolak2004}), we use a temperature of 100 K in all models to provide a consistent reference for comparing the relationship between luminosity and the snowline in models of different density structures.

Figure \ref{fig:fig04} shows that the water snowline is located at different radii as a result of the different density structures. The envelope-only Model E shows water snowlines located at 40 au and 150 au for luminosities of 10 $L_{\odot}$ and 100 $L_{\odot}$, respectively. However, Model E+D and PPD with disk structure have water snowlines located several times closer (16 au and 48 au for Model E+D, and 5 au and 16 au for Model PPD). This implies that the relation between luminosity and water snowline varies depending on the density structure.

Figure~\ref{fig:fig05} and Table \ref{tab:table1} show the relations between luminosity and the water snowline in the models. The relations derived for the low-mass YSO models follow the form $R_{\rm snow}=a\times (L/L_{\odot})^p$ au, where the factor $a$ varies depending on density structure. For Model E, an increase in the envelope mass to 0.5 to 6.5 $M_{\odot}$ leads to the water snowline forming farther from the center at a given luminosity, increasing the factor $a$ in relation to 12.9–19.4 au. The power index $p$ ranges from 0.49 to 0.45 as the envelope mass increases.

\citet{vant2022} reported that the water snowlines in various envelope models have a 20–30\% uncertainty relative to the prediction based on Equation \ref{eq:bisschop} from \citet{Bisschop2007}, which uses a factor $a$ = 15.4 and power index $p$ = 0.5.  The values of $a$ and $p$ in our envelope-only models with different envelope masses are consistent with the uncertainty range of \citet{vant2022}. It was also noted that larger envelope masses tend to confine radiation on smaller scales, shifting the water snowlines outward compared to lower-mass envelopes. Our envelope-only models also show that the snowline moves outward with increasing envelope mass, consistent with the model result reported in \citet{vant2022}.

For the relation of Model E+D, the factor $a$ is derived to be 4.46 au (Figure \ref{fig:fig05}, Table \ref{tab:table1}). This value is $\sim$3 times lower than that of Model E. The power index $p$ is derived to be 0.51, which is consistent with Model E. The smaller value of the factor $a$ in Model E+D can be explained by the presence of the disk structure. Figure \ref{fig:fig06} presents the dust density and temperature profiles at 100 $L_{\odot}$ of each model in the midplane. Model E has a water snowline at 150 au, and Model E+D has a water snowline at 48 au. These two models share the same envelope density profile. However, Model E+D has a disk with a size of 90 au, which is significantly denser than the envelope. This dense disk structure blocks radiation emitted from the central protostar, resulting in lower dust temperatures compared to Model E. Thus, the disk structure restricts the water snowline to a smaller radius, which leads to a smaller value of the factor $a$.

\citet{Murillo2022} also found that when the protostellar system has a disk structure, the water snowline is limited within the disk, based on radiative transfer models. In addition, they reported that for the high luminosity and small disk, the water snowline is located in the envelope. Consistently, in our model, the water snowlines in Model E+D are located in the envelope, at luminosities above 215 $L_{\odot}$ (Figure \ref{fig:fig05}). However, due to the disk structure, the envelope of Model E+D is still colder than that of Model E (Figure \ref{fig:fig06}), leading to a significantly closer water snowline and a lower value of the factor $a$.

The blocking effect of central radiation by the disk is enhanced in Model PPD. The disk of Model PPD is much denser than in Model E+D, leading to the water snowline at a much closer distance of 20 au (Figure \ref{fig:fig06}). Therefore, the factor $a$ in the relation is derived to be 1.45, which is $\sim$10 times lower than that of Model E. The power index $p$ is derived to be 0.54, which is also consistent with Model E.

\begin{table}[t!]
\caption{Relation between luminosity and water snowline\label{tab:table1}}
\centering
\textbf{Relation:} $R_{\rm snow}=a\times (L/L_{\odot})^p$ [au] \\
\centering
\begin{tabular}{lrr}
\toprule
Model                                      & $a$& $p$\\
\midrule
Model E ($M_{env}$ = 0.5 $M_{\odot}$)&  12.9&  0.49\\
Model E ($M_{env}$ = 1.0 $M_{\odot}$)&  13.2& 0.49\\
Model E ($M_{env}$ = 2.0 $M_{\odot}$)&  14.3&  0.48\\
Model E ($M_{env}$ = 5.0 $M_{\odot}$)&  18.0&  0.46\\ 
Model E ($M_{env}$ = 6.5 $M_{\odot}$)&  19.4&  0.45\\ \addlinespace
Model E+D&  4.46&  0.51\\ \addlinespace
Model PPD (without viscous heating)&  1.45&  0.54\\
Model PPD (with viscous heating)&  4.97&  0.37\\
\bottomrule
\end{tabular}
\end{table}

\begin{figure*}[t]
\centering
\includegraphics[width=\textwidth]{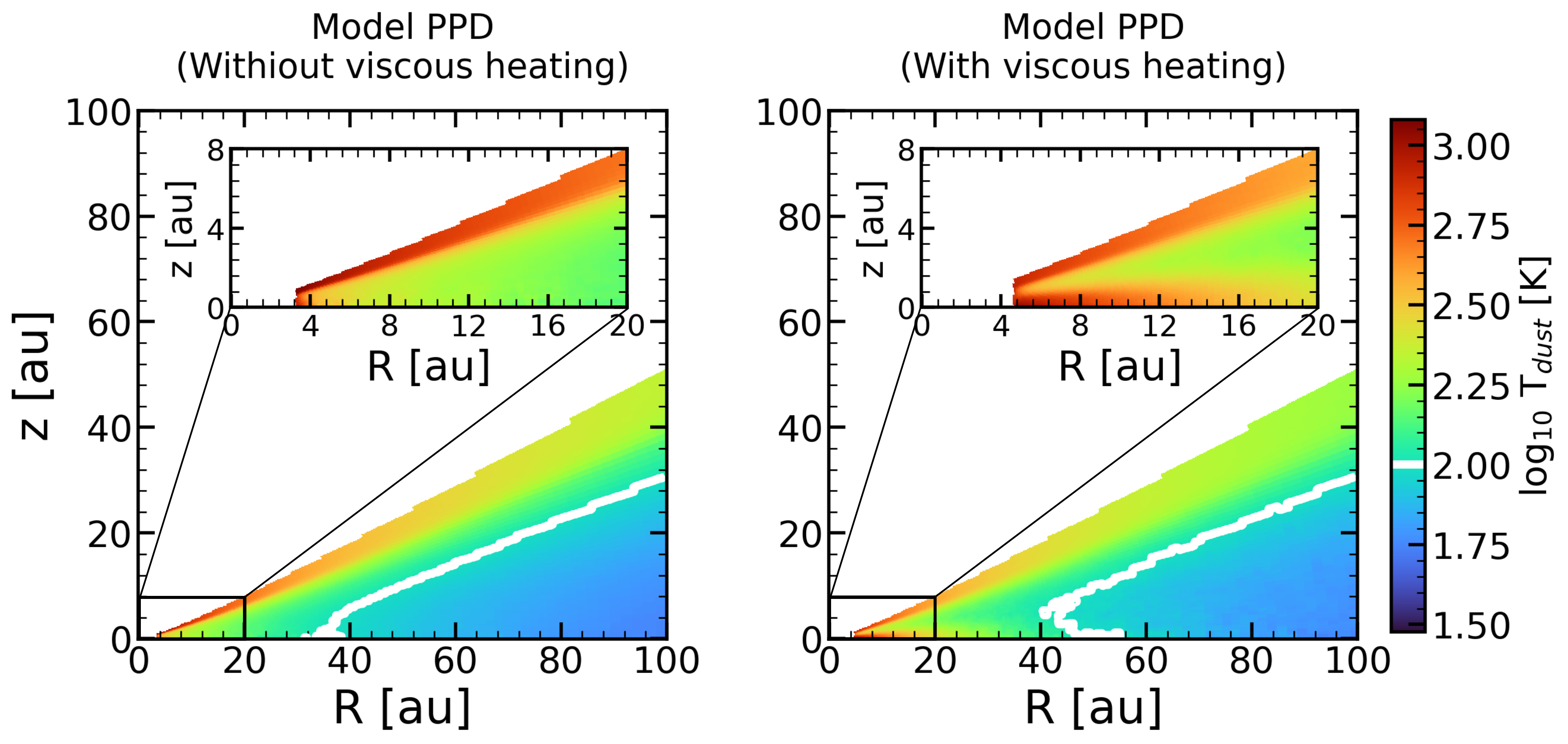}
\caption{The 2-dimensional dust temperature distributions in Model PPD at the total luminosity of $L_{\rm tot} =$ 400 $L_{\odot}$ (fiducial model) without viscous heating \textit{(left)} and with viscous heating \textit{(right)}. The white solid lines indicate the isothermal contour of $T_{\rm dust}$ = 100 K, and the inset in the upper panels zooms in on the inner 20 au$\times$8 au.\label{fig:fig07}}
\end{figure*}

\begin{figure}[t]
\centering
\includegraphics[width=0.45\textwidth]{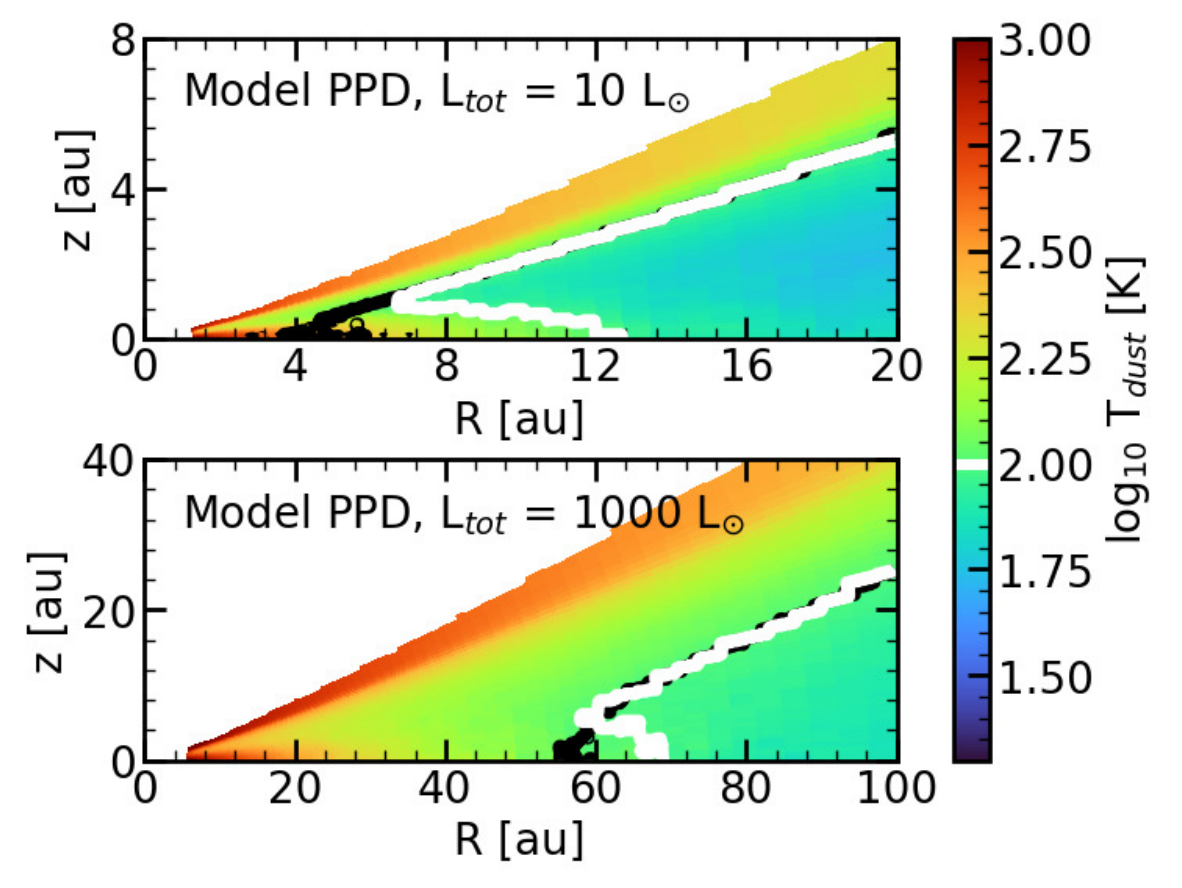}
\caption{The 2-dimensional dust temperature distributions in Model PPD at the total luminosity of $L_{\rm tot}$ = 10 $L_{\odot}$ (\textit{top}) and $L_{\rm tot}$ = 1,000 $L_{\odot}$ (\textit{bottom}) including viscous heating. The white solid line indicates the isothermal contour of $T_{\rm dust}$ = 100 K with viscous heating, and the black solid line also represents $T_{\rm dust}$ = 100 K without viscous heating.\label{fig:fig08}}
\end{figure}

Thus, depending on the density structure of low-mass YSOs, the presence of a disk and increasing disk density lead to a decrease in the factor $a$, reproducing a closer water snowline in the protoplanetary disk at the same luminosity. We note that the value of factor $a$ and power index $p$ in the relation may differ depending on the disk properties, and our results present the trend of the water snowline and its relation according to the various density structures.

We also explore the effect of viscous heating on the water snowline. Figure \ref{fig:fig07} shows that for the fiducial model of Model PPD ($L_{\rm tot} =$ 400 $L_{\odot}$), dust grains in the midplane within 20 au are heated to a few hundred Kelvin due to viscous heating. This causes the water snowline to shift outward from 36 au to 50 au, despite the same total luminosity. Figure \ref{fig:fig05} also shows that adopting viscous heating in Model PPD with $L_{\rm tot}$= 10–1,000 $L_{\odot}$ shifts the water snowline by a few to 15 au, modifying the relation of the water snowline accordingly. In the relation, the factor $a$ of Model PPD with viscous heating is derived to be 4.97 au, which is 3.4 times higher than 1.46 au of the models without viscous heating. However, the power index $p$ is derived as 0.37, which is significantly lower than 0.5 (Table \ref{tab:table1}), implying that the location of the snowline extends slowly compared to models without viscous heating.

The relations could be modified depending on the importance of the viscous heating at a given total luminosity. Figure \ref{fig:fig08} shows the effect of viscous heating on the water snowline of Model PPD at  $L_{\rm tot}$= 10 and 1,000 $L_{\odot}$. In the case of the $L_{\rm tot}$= 10 $L_{\odot}$ model, viscous heating acts as a dominant heat source at 4 au, which corresponds to the water snowline in the model without viscous heating. Thus, at low luminosity, viscous heating plays an important role in determining the location of the water snowline, and the shifted snowline of 12 au is located significantly farther away compared to the 4 au snowline. This causes the factor $a$ to increase from 1.46 au to 4.97 au. 

However, for the $L_{\rm tot}$= 1,000 $L_{\odot}$ model, the water snowline in the model without viscous heating (56 au) is significantly extended to a large radius. Despite adopting viscous heating, the water snowline shows no significant change at high luminosity. This means that the water snowline at high luminosity is primarily determined by irradiative (passive) heating. This causes the power index $p$ to decrease from 0.5 to 0.37. We need to explore whether the effect of viscous heating on the water snowline at each luminosity shows the same trend across different disk density structures. In Section \ref{sec:discussion_2}, we discuss how viscous heating affects the water snowline in Model E+D and PPD under the same total luminosity.


\begin{figure*}[t]
\centering
\includegraphics[width=\textwidth]{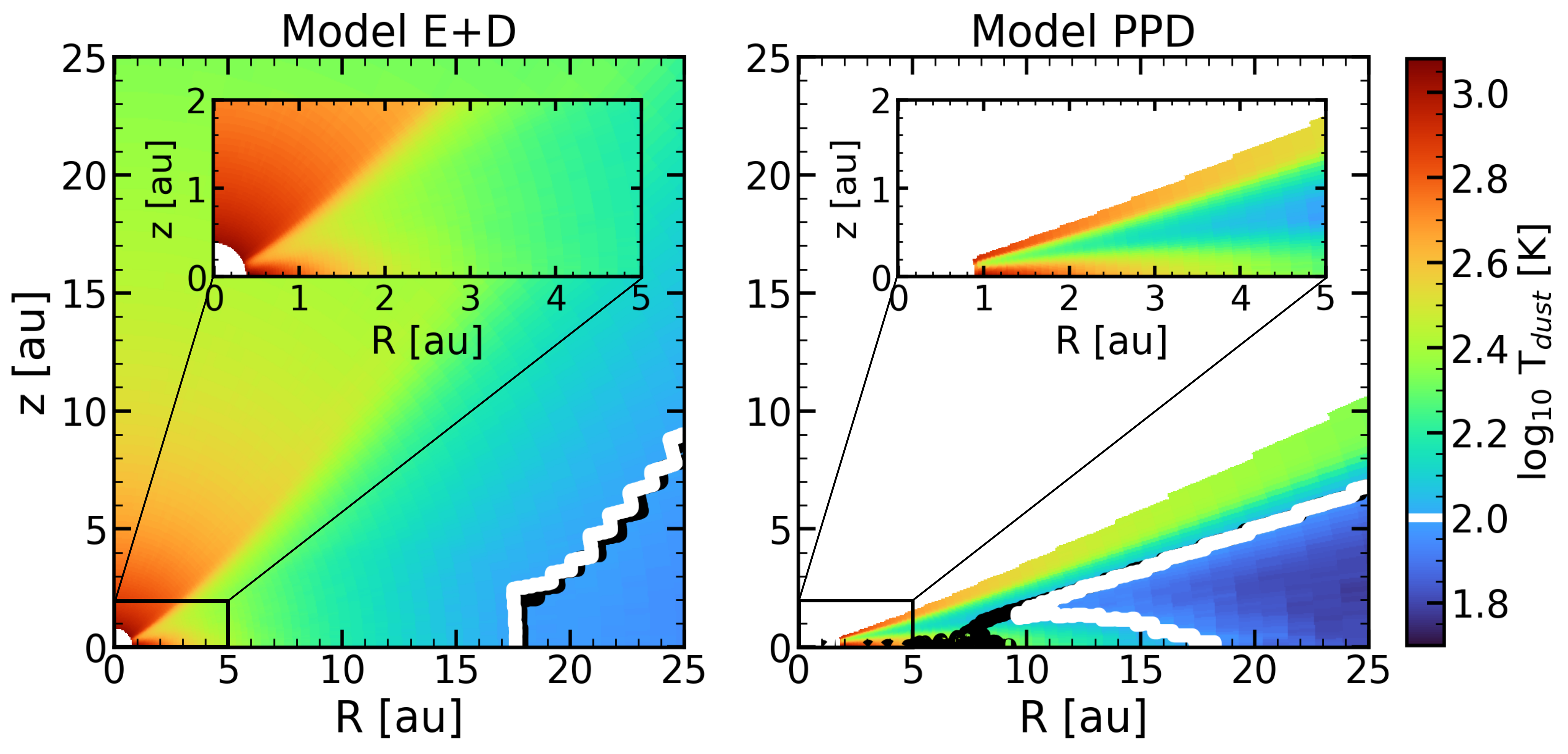}
\caption{The 2-dimensional dust temperature distribution at $L_{\rm tot}$ = 22 $L_{\odot}$ including viscous heating. Model E+D (\textit{left}) and Model PPD (\textit{right}) are presented. The white solid line indicates the isothermal contour of $T_{\rm dust}$ = 100 K with viscous heating, and the black solid line also represents $T_{\rm dust}$ = 100 K without viscous heating. The inset in both upper panels zooms in on the inner 5 au$\times$2 au.}\label{fig:fig09}
\end{figure*}

\section{Discussion}
\label{sec:discussion}
\subsection{The effect of disk structure in Class 0 stage}
\label{sec:discussion_1}
B335 is a Class 0 YSO for which the water snowline has been estimated using various molecular emissions (\citealt{Jensen2019,Jensen2021,jelee2025}). Based on WISE / NEOWISE observation in the W2 band, the luminosity was estimated to be 3 $L_{\odot}$ in the quiescent phase \citep{Evans2023}. During this period, due to the low luminosity and angular resolution limitations, the water snowline could not be observed.

Later, \citet{Evans2023} and \citet{Kim2024} reported that B335 had undergone an outburst, increasing to a maximum luminosity of 22 $L_{\odot}$. During the burst, the water snowline was extended and became observable due to the increased luminosity. \citet{Jensen2019,Jensen2021} observed the water isotopologue lines, estimating the water snowline to be 10–14 au \citep{vant2022}. In addition, since COMs are sublimated with water ice (\citealt{Herbst2009, vanDishoeck2013}), \citet{jelee2025} used methanol emission, a well-known tracer of the water snowline (\citealt{vant2018b, Lee2019}), to predict that the water snowline in B335 is located at $\sim$20 au.

However, the envelope-only model (\citealt{Bisschop2007, vant2022}) estimates the water snowline to be 72 au in the burst phase (22 $L_{\odot}$), and it is several times farther than the observations (10–20 au). Furthermore, the model predicts that even during the quiescent phase (3 $L_{\odot}$), the water snowline is located at 25 au, which is the observable radius. This discrepancy between the observed water snowlines and those predicted by the envelope-only model suggests the possibility of a dense disk-like structure within the protostellar envelope.

Our Model E+D predicts the water snowline to be located at 8 au and 21 au for luminosities of 3 $L_{\odot}$ and 22 $L_{\odot}$, respectively (Figure \ref{fig:fig05}, Table \ref{tab:table1}). The dense disk structure blocks radiation from the center, resulting in snowlines having radii that are several times smaller than those predicted by the envelope-only model. As a result, the model reproduces a water snowline at a few au, which is difficult to observe during the quiescent phase. In addition, during the burst phase, the model prediction (21 au) better reproduces the snowline traced by molecular lines (10–20 au).

The presence of a disk structure in B335 has also been inferred from dust continuum observation and radiative transfer modeling. \citet{Yen2015b} observed a dense disk-like structure with a radius of less than 16 au using 1.3 mm dust continuum emission, and \citet{Evans2023} suggested that the disk structure should be included in a radiative transfer model to reproduce the observed SED.

Thus, by comparing our model relation with the observed water snowline in Class 0 protostellar systems, the discrepancy of the water snowline can serve as a tracer of the disk structure. This suggests that, even for protostellar envelopes, a relation that accounts for the presence of a disk may be necessary when tracing the water snowline.

\subsection{The effect of viscous heating}
\label{sec:discussion_2}
V883 Ori is currently undergoing a burst phase with active viscous heating. The current luminosity of V883 Ori is 218–647 $L_{\odot}$ (\citealt{Strom1993, Furlan2016, Liu2022}) and the water snowline was estimated to be 42–80 au in the dust continuum and in the water isotopologue observations (\citealt{Cieza2016, Lee2019, Tobin2023, Lee2024}).

The snowline predicted by our PPD model without viscosity heating (26–43 au; Figure \ref{fig:fig05}; Table \ref{tab:table1}) better reproduces the observed water snowline (42–80 au) than that derived from the envelope-only model (230–360 au; \citealt{Bisschop2007, vant2022}). However, this prediction is $\sim$2 times smaller than the observed snowline. In Figure \ref{fig:fig07}, we show that applying viscous heating shifts the water snowline outward. Thus, our viscous heating protoplanetary disk model (Figure \ref{fig:fig05}, Table \ref{tab:table1}) predicts the water snowline as 36–55 au in the luminosity range of 218–647 $L_{\odot}$ and better reproduces the observed water snowline.

\citet{Alarcón2024} adopted viscous heating in the radiative transfer model to explain the thermal structure of the disk of V883 Ori. They showed that the water snowline ($T_{\rm dust}$=115K) in V883 Ori is shifted from 20 au to 30 au when viscous heating is included in the model at 400 $L_{\odot}$. Our fiducial model of Model PPD (400 $L_{\odot}$; Figure \ref{fig:fig07}), based on V883 Ori, also shows that the water snowline ($T_{\rm dust}$=100K) shifts outward from 36 au to 50 au with consideration of viscous heating. For a higher water sublimation temperature of 115 K, the snowline in our model shifts from 30 au to 40 au when viscous heating is taken into account. This implies that applying viscous heating to a protoplanetary disk model, such as V883 Ori, affects the prediction of the water snowline.

We also explore the influence of viscous heating in Model E+D. In Figure \ref{fig:fig09}, Model PPD shows a 10 au shift in the water snowline at $L_{\rm tot}$ = 22 $L_{\odot}$ if viscous heating is included. In contrast, there is no significant shift when viscous heating is considered in Model E+D. In the case of Model PPD with viscous heating at $L_{\rm tot}$ = 22 $L_{\odot}$, viscous heating is the dominant heating source in the disk midplane. However, for Model E+D, the viscously heated disk appears only at very small radii in the midplane at the same total luminosity. This means that Model E+D has weak viscous (active) heating, so irradiative (passive) heating becomes the main heating source, resulting in the effect of viscous heating on the water snowline being negligible.

The importance of viscous heating depends on the disk mass. \citet{Takakuwa2024} found that the radiative transfer model considering only passive heating cannot explain the high brightness temperature ($\sim$195 K) observed in the 1.3 mm dust continuum in the Class I YSO R CrA IRS7B-a. According to their work, viscous heating is crucial for reproducing the high brightness temperature. The disk of R CrA IRS7B-a ($M_{\rm disk}$= 0.41 $M_{\odot}$) is much more massive than our Model E+D ($M_{\rm disk}$= 0.0075 $M_{\odot}$), and viscous heating is expected to influence the hot thermal structure in the disk strongly. This is also expected to shift the water snowline outward.

Consistently, since Model PPD ($M_{\rm disk}$= 0.23 $M_{\odot}$) has a more massive disk than Model E+D, viscous heating becomes the dominant heating source in the midplane, resulting in a significant shift of the water snowline. Thus, it should be noted that the effect of viscous heating on the water snowline can vary depending on the disk mass.

\subsection{An indicator of burst events in YSOs}
\label{sec:discussion_3}
Since our models trace the water snowline based on dust temperature, the predictions may differ from the actual location. However, by comparing our model predictions with observed water snowlines, we can offer insight into recent burst events in YSOs. This is because, although dust grains cool instantaneously after an accretion burst ends, sublimated molecules can remain in the gas phase since their freeze-out timescale onto dust grains is longer than the cooling timescale of the dust grains (\citealt{Lee2004,Lee2007}). The water freeze-out timescale is 100-1,000 years (\citealt{Visser2012,Visser2015}) in the protostellar envelope.

If a protostar experienced a burst accretion event and water or COM lines were observed before water molecules re-freeze onto grain surfaces, the measured water snowline could appear at a radius larger than that predicted by the dust temperature. \citet{vant2022} demonstrated that for the Class 0 protostar IRAS 15398, water molecular line observations reveal a snowline five times farther out than predicted based on the current luminosity. This discrepancy between the snowline measured from molecular emission and that estimated from the current luminosity can be used to trace recent outburst events in YSOs.

\section{Conclusion}
\label{sec:conclusion}
To explore the water snowline in various density structures of low-mass YSOs, we construct dust continuum radiative transfer models for an envelope-only model (Model E), an envelope + disk + cavity model (Model E+D) and a protoplanetary disk model (Model PPD), and trace the water snowline in a luminosity range of 0.1–1,000 $L_{\odot}$.

\vspace{3mm}
\begingroup
\leftskip=5mm
\noindent 1. In all low-mass YSO models, the luminosity and the water snowline follow the relation $R_{\rm snow}=a\times (L/L_{\odot})^p$ au. In this relation, the factor $a$ decreases from Model E to PPD because the denser disk structure blocks the central radiation, confining the water snowline to a smaller radius at the same total luminosity. However, the power index $p$ consistently retains a value of $\sim$0.5.
\vspace{3mm}
\par
\endgroup

\begingroup
\leftskip=5mm
\noindent 2. Viscous heating in Model PPD shifts the water snowline outward by a few to 15 au for a given luminosity range, leading to an increase in the factor $a$ and a decrease in the power index $p$ for the relation. However, the effect of viscous heating is negligible in Model E+D, resulting in no significant shift in the location of the water snowline. This implies that the effect of viscous heating on the water snowline depends on the mass of the disk.
\vspace{3mm}
\par
\endgroup

\begingroup
\leftskip=5mm
\noindent 3. The water snowline of V883 Ori (42–80 au), which is currently in the burst to be 218–647 $L_{\odot}$, is better reproduced by Model PPD (26–43 au, without viscous heating) compared to the prediction of previous envelope-only models (230–360 au). The $\sim$2 times of discrepancy between the Model PPD prediction and observation decreases by applying viscous heating, yielding the water snowine in a range of 36–55 au.
\vspace{3mm}
\par
\endgroup

\begingroup
\leftskip=5mm
\noindent 4. Our model traces the water snowline using dust temperature, and therefore has the limitation of not accounting for line observations. Nevertheless, comparing our models with molecular line observations allows us to infer the bursting events and disk structure in low-mass YSOs.
\par
\endgroup


\acknowledgments
This research was funded by the New Faculty Startup Fund of Seoul National University and by the National Research Foundation of Korea (NRF) through grants provided by the Korean government (MSIT) (grant numbers 2021R1A2C1011718 and RS-202400416859). GB was supported by Basic Science Research Program through the National Research Foundation of Korea (NRF) funded by the Ministry of Education (RS-2023-00247790).


{}


\end{document}